\documentclass[aps,prb,amsfonts,amssymb,twocolumn,showpacs,
superscriptaddress,footinbib]{revtex4}

\usepackage{graphicx}
\usepackage{amsfonts}
\usepackage{amssymb}
\usepackage{amsmath}
\usepackage{verbatim}

\usepackage{flafter}  

\begin{document}

\title{Quantum instability in a dc-SQUID with strongly asymmetric dynamical 
parameters}

\author{A.U. Thomann}
\affiliation{Institute for Theoretical Physics, ETH Zurich, 8093 Zurich, Switzerland}
\author{V.B. Geshkenbein}
\affiliation{Institute for Theoretical Physics, ETH Zurich, 8093 Zurich, Switzerland}
\affiliation{L.D. Landau Institute for Theoretical Physics, 117940 Moscow, Russia}
\author{G. Blatter}
\affiliation{Institute for Theoretical Physics, ETH Zurich, 8093 Zurich, Switzerland}



\date{\today}

%
\begin{abstract} 
A classical system cannot escape out of a metastable state at zero
temperature.  However, a composite system made from both classical and quantum
degrees of freedom may drag itself out of the metastable state by a sequential
process. The sequence starts with the tunneling of the quantum component which
then triggers a distortion of the trapping potential holding the classical
part. Provided this distortion is large enough to turn the metastable state
into an unstable one, the classical component can escape. This process reminds
of the famous baron M\"unchhausen who told the story of rescuing himself from
sinking in a swamp by pulling himself up by his own hair---we thus term this
decay the `M\"unchhausen effect'.  We show that such a composite system can be
conveniently studied and implemented in a dc-SQUID featuring asymmetric
dynamical parameters.  We determine the dynamical phase diagram of this system
for various choices of junction parameters and system preparations.
\end{abstract}
%

\maketitle

%
\section{Introduction}
\label{sec:introduction}
%

Consider a classical object (a degree of freedom) trapped in a metastable
potential minimum; no decay out of this metastable state is possible at low
temperatures, where thermal activation over the barrier is exponentially
suppressed. However, if the classical object is a composite one, with a
quantum degree of freedom coupled to the classical one, then the quantum
object may tunnel out of the metastable minimum and exert a pulling force on
the classical object. Once the latter is large enough to completely suppress
the trapping barrier, the classical object is able to leave the potential 
well---hence a classical object may escape from a metastable state even at 
zero temperature if helped by a coupled quantum degree of freedom.

\begin{figure}[bht]
\includegraphics{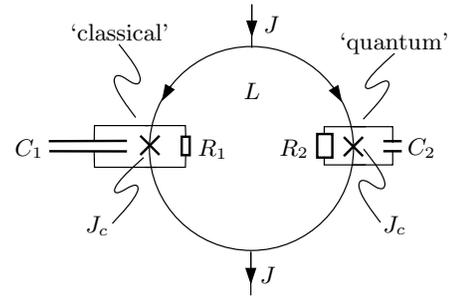}
\centering
\caption{Schematics of the dynamically asymmetric dc-SQUID. Two Josephson
junctions are integrated in a current ($J$) biased superconducting loop with
inductance $L$. The two junctions feature equal critical currents
$J_\mathrm{c}$ but strongly asymmetric (shunt) capacitances $C_i$ and (shunt)
resistances $R_i $. We assume that $C_i$ and $R_i$ are chosen such that
quantum effects are present for junction 2 but are strongly suppressed for
junction 1, hence $C_1 \gg C_2$ and/or $R_1 \ll R_2$.} \label{fig:setup}
\end{figure}

The above situation can be realized in a dc-SQUID (Superconducting Quantum
Interference Device) featuring asymmetric dynamical parameters; i.e., with two
Josephson junctions of equal critical currents $J_\mathrm{c}$ but strongly
different (shunt) capacitances $C$ and (shunt) resistances $R$, see 
Fig.\ \ref{fig:setup}; choosing large and small parameters $C$ and $1/R $ for the
two junctions allows to place one of the junctions in the `classical' and the
other into the quantum domain. The tunneling of the quantum degree of freedom
entails a distortion of the trapping potential of the classical junction,
which might be sufficiently large to transform the metastable state of the
classical junction into an unstable one.  The appearance of this complex decay
channel depends critically on the applied bias current $J$ and the SQUID's
loop inductance $L$ coupling the two junctions.

The gauge invariant phase differences $\varphi_i$, $i=1,2$, across the two
Josephson junctions\cite{Josephson:1962} define our dynamical degrees of
freedom: Assuming equal critical currents $J_c$, the junctions' potential
energies ${\cal V}_i =E_J [1-\cos \varphi_i]$, $i=1,2$, involve the Josephson
energy $E_J =\Phi_0 J_c/2\pi c$ (with $\Phi_0 = h c/2e$ the flux unit, $e$ and
$c$ denote the unit charge and light velocity).\cite{footnote1} Their
kinetic energies ${\cal T}_i = (\hbar/2e)^2 C_i \dot\varphi_i^2 /2$ are
determined by the junction capacitances $C_i$ playing the role of effective
masses (the relevant energy scale is given by the charging energy
$E_{c,i}=e^2/2C_i$)---a dynamically asymmetric SQUID with one large and one
small junction capacitance then provides us with the desired classical and
quantum degrees of freedom (we choose $C_1 \gg C_2$; additional normal
resistances $R_i$ introduce a dissipative dynamical component, see below). The
coupling of the two junctions via the loop inductance $L$ produces the
interaction energy ${\cal V}_\mathrm{int} = [(\Phi_0/2\pi)^2/2 L]
(\varphi_2-\varphi_1)^2$ involving the relative coordinate
$(\varphi_2-\varphi_1)$, whereas the external driving current $J$ couples to
the absolute coordinate, ${\cal V}_{\rm drive} = E_J (J/2J_c)
(\varphi_1+\varphi_2)$.  While large-capacitance (`classical') junctions are
easily fabricated, small (`quantum') junctions are more difficult to realize.
Nevertheless, experimental techniques to fabricate small junctions are
available today and their quantum behavior in the form of quantum tunneling,
\cite{Voss:1981, Caldeira:1981, Caldeira:1983, Devoret:1985} quantized energy
levels \cite{Martinis:1985, Martinis:1987} and even quantum coherence
\cite{Chakravarty:1984, Leggett:1987, Nakamura:1997, Nakamura:1999,
Friedman:2000, Wal:2000, Chiorescu:2003} has been demonstrated.

The quantum decay of the biased, dynamically symmetric dc-SQUID has been
discussed before both experimentally\cite{Li:2002,Balestro:2003} and
theoretically,\cite{Chen:1986} also in the context of instanton
splitting.\cite{Ivlev:1987, Morais-Smith:1994} In the SQUID discussed here,
\cite{footnote2} the large
asymmetry of the dynamical parameters blocks the tunneling of the $\varphi_1$
degree of freedom; the `heavy' junction then remains frozen during the quantum
decay of the `light' degree of freedom $\varphi_2$.  A trajectory of this kind
describes the entry of flux into the SQUID loop which rearranges the current
flow in the two arms in a way as to redirect more current through the heavy
junction. This increase in current produces an enhanced tilt $-J_\mathrm{eff}
\varphi_1$ in the potential of the heavy junction, which then may decay
through a classical trajectory, see Fig.\ \ref{fig:corrpot}. We call this
nontrivial decay sequence the `{M\"unchhausen decay}'.  It is the aim of this
work to determine the effective critical current $J_c(L^{-1})$ for which the
M\"unchhausen decay becomes possible, see  Figs.\ \ref{fig:phase_diagram_vv}
and \ref{fig:phase_diagram_adiabatic}-\ref{fig:pe}.

In the following, we define our system in full detail, including a dissipative
component in the junction dynamics (Sec.\ \ref{sec:setup}). Sections\
\ref{sec:overdamped} and \ref{sec:underdamped} are devoted to the derivation
of the effective critical currents for the various cases with junctions
governed by massive or dissipative dynamics. In Section\ \ref{sec:remarks} we
add remarks concerning the experimental realization of the system described
here. Finally, we draw conclusions in Section\ \ref{sec:conclusion}.

\section{Setup and model}
\label{sec:setup}

\begin{figure}[htb]
\centering
\includegraphics{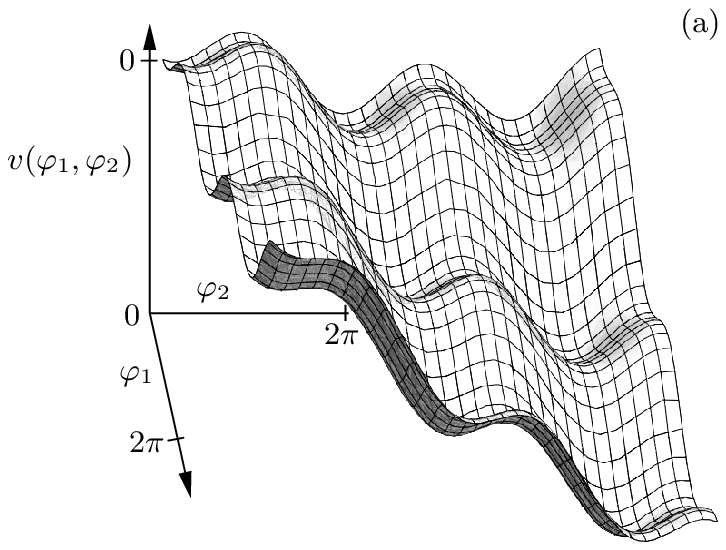}%
\linebreak
\includegraphics{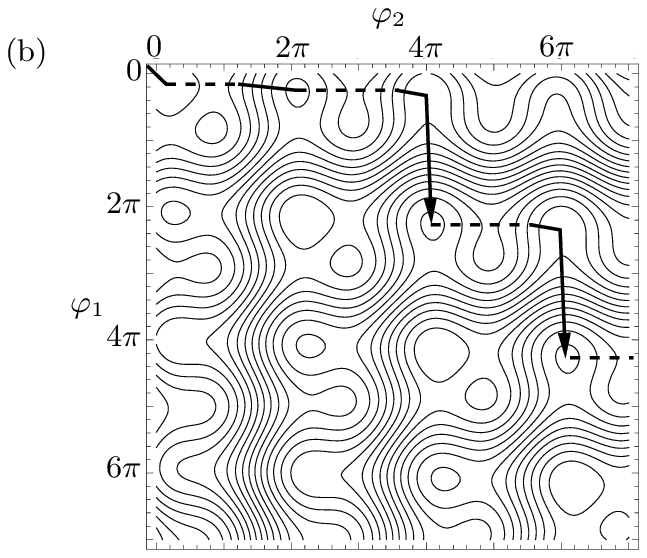}%
\caption{Surface (a) and contour plot (b) of the SQUID potential, Eq.\
\eqref{eq:v}, at bias current $j=0.5$ and coupling $k=0.04$.  For an
undercritical current $j<1$, the symmetric minima  $(\varphi_1=\varphi_2)$ are
all metastable. The stability of the side-minima  $(\varphi_1\neq \varphi_2)$
depends crucially on the parameters $j$ and  $k$. The line in (b) shows the
decay path in an overdamped setup: The system starts out in a relaxed local
ground state near the initial minimum at $\varphi_1=\varphi_2=\arcsin j$. The
`light' degree of freedom $ \varphi_2$ then tunnels (dotted line) and the
system relaxes to the bottom of the next minimum near $\varphi_2\approx 2\pi$,
with a classically stable but quantum mechanically metastable ground state.
Through an additional quantum phase slip, the light phase reaches the minimum
near $\varphi_2\approx 4\pi$, which is \emph{not} classically stable and
henceforth the system can decay along a classically allowed path to
$\varphi_1\approx 2\pi$. The system then has turned unstable and enters a
resistive state through iteration of the last two steps.}
\label{fig:corrpot} 
\end{figure}

Within the resistively and capacitively shunted junction (RCSJ) model (at $T=0$), the 
classical dynamics of the two phase differences $\varphi_1$ and $\varphi_2$ is 
governed by the equations of motion
\begin{equation}
  \omega_{0,i}^{-2} \ddot{\varphi}_i +\eta_i \dot{\varphi}_i
  = - \partial_{\varphi_i}v(\varphi_1,\varphi_2),
  \label{eq:motion}
\end{equation}
with the plasma frequency $\hbar^2\omega_{0,i}^2=8 E_J E_{c,i}$ of an 
unbiased single junction and the damping coefficients $\eta_i = \Phi_0 / 2 \pi c 
J_c R_i$ ($R_i$ denote the normal ohmic junction resistances). The 
potential (see Fig.\ \ref{fig:corrpot}) is given by
\begin{multline}
        v(\varphi_1,\varphi_2)=1-\cos\varphi_1+1-\cos\varphi_2 \\
        -j (\varphi_1+\varphi_2)+\frac{k}{2} (\varphi_1-\varphi_2)^2,
        \label{eq:v}
\end{multline}
with the dimensionless current $j = J/2J_c$ and the coupling constant $k=
\Phi_0 c / 2\pi J_c L = 1/\beta_L$ ($\beta_L$ denotes the usual screening
parameter of the SQUID). Eqns.\ \eqref{eq:motion} and \eqref{eq:v} describe a
dc-SQUID with symmetric inductance $L$ in a vanishing external magnetic field
and driven by a bias current $J$ or, equivalently, the massive (mass $\propto
\omega_{0,i}^{-2}$) and/or dissipative ($\eta_i$) dynamics of two harmonically
($k$) coupled particles in a tilted ($j$) and corrugated ($\cos \varphi_i$)
potential. Quantum effects of the light junction 2 are accounted for via the
relevant tunneling and decay processes, see below.

\begin{figure}[htb]
\includegraphics{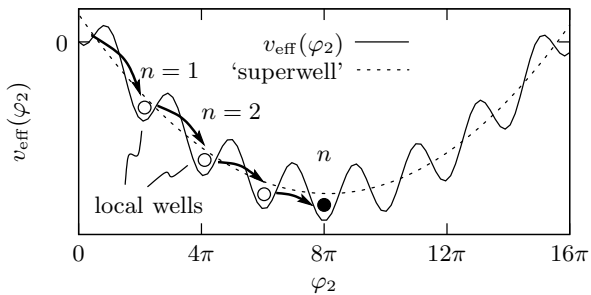}
\caption{Effective potential $v_\mathrm{eff}(\varphi_{2})=v(\varphi_1
=\mathrm{const.}, \varphi_2)$ (solid line) and the parabola remaining after
dropping $\cos\varphi_2$ (dashed line) for $j=0.5$, $k=0.02$ and
$\varphi_1=\arcsin j$. The bullets and arrows illustrate the sequential decay
of the quantum phase $\varphi_2$ to the ground state for the case of strong
damping.}
\label{fig:effpot} 
\end{figure}

The potential \eqref{eq:v} gives rise to two types of relevant frequencies.
One is the plasma frequency $\omega_{p,i}$, the small-amplitude frequency in
the direction of $\varphi_i$ around a local minimum of the potential
$v(\varphi_1,\varphi_2)$. With the effective potential
\begin{equation}
   v_\mathrm{eff}[\varphi_l](\varphi_i)
   =v(\varphi_l=\mathrm{const.},\varphi_i),\ i\neq l,
\label{eq:effpot}
\end{equation}
cf.\ Fig.\ \ref{fig:effpot}, $\omega_{p,i}^2 = \omega_{0,i}^2
\partial_{\varphi_i}^2 v_ \mathrm{eff}(\varphi_i)$, evaluated at a local
minimum $\varphi_i^\mathrm{min}$, and depends on the parameters $j$ and $k$ 
as well as on $\varphi_i^\mathrm{min}$.  For the heavy junction (junction 1),
$\omega_{p,1}$ can become arbitrarily small upon approaching criticality,
while for the quantum junction (junction 2) $\omega_{p,2}$ becomes small only
for $j\rightarrow 1$ and $k \to 0$.  
The other frequency is given by the $LC$
constant of the `superwell' in $v_ \mathrm{eff}(\varphi_2)$, cf.\ Fig.\
\ref{fig:effpot}, and is relevant only in the regime $k\ll 1$ and for the
quantum junction (junction 2), 
$\omega_\mathrm{LC,2}^2 = \omega_{0,2}^2 k=c^2/LC_2$.

These characteristic frequencies delineate the regimes where the two junctions
behave classically and quantum mechanically, respectively. The relevant
physical parameters are given by the ratios $E_J/E_{c,i}$.  For the heavy
junction, we have to make sure that the number $N_l \approx E_b/\hbar
\omega_{p,1} \sim (E_J/E_{c,1})(1-J_1/J_c)^{5/4}$ of states in the local well
remains large.  Choosing a large ratio $E_J/E_{c,1} \gg 1$ then guarantees 
that quantum effects (i.e., $N_l$ of order unity) become relevant only very
close to criticality and we choose to ignore them in our discussion below.

The system develops the most interesting behavior if the quantum degree of
freedom develops well localized states within all relevant local minima in $v_
\mathrm{eff}(\varphi_2)$, Fig.\ \ref{fig:effpot}, and hence parameters should
be chosen such that the quasi-classical description applies. On the other
hand, we require tunneling and coherence effects to manifest themselves on
reasonably short (measurable) timescales. We then choose a ratio
$E_J/\hbar\omega_{0,2}\sim 2$, such that the local wells in $v_\mathrm{eff}
(\varphi_2)$ contain a few quasi-classical states each.

The strength of dissipation can be quantified by the dimensionless damping 
parameters, 
\begin{eqnarray}
\alpha_{p,i}&=&(2 R_i C_i \omega_{p,i})^{-1},\\
\alpha_{LC,2}&=&(2 R_2 C_2 \omega_{LC,2})^{-1}.
\end{eqnarray}
Below, we are interested in the two limiting cases of strong and weak damping.
For a strongly damped quantum junction with $\alpha_{p,2}>1$, 
the quantum decay of $\varphi_2$ out of a metastable well of
$v_\mathrm{eff}(\varphi_2)$ is incoherent \cite{Caldeira:1983,Leggett:1987}
and its subsequent relaxation is fast (as compared to the dynamics of
$\varphi_1$).  For weak damping, $\alpha_{LC,2}\ll1$, $\alpha_{p,i}\ll1$, the
kinetic energy stored in the motion of the heavy junction has to be accounted
for; in addition, the finite lifetime of the quantum states of the light
junction due to the residual dissipation has to be considered, see the
discussion in Sec.\ \ref{sec:underdamped}.

\section{Strong damping}
\label{sec:overdamped}

We start by analyzing the situation for strong damping, $\alpha_{p,2}>1$ and
$\alpha_{p,1}\gg 1$, see Ref.\ \onlinecite{Thomann:2008} for an earlier short
report. We bear in mind an experiment with a dc-SQUID characterized by a fixed
inductance $L\propto k^{-1}$ and biased with a current $j < 1$. The task is to
determine whether the M\"unchhausen decay can take place; in the experiment,
the latter manifests itself through the transition to a finite voltage state.
In the strong damping case, no kinetic energy is stored in the system.
Furthermore, the evolution is not sensitive to the way the current is ramped.
After current ramping, the system starts out in a relaxed state where the
phases $\varphi_i$ are localized in the diagonal metastable minimum at
$\varphi_1=\varphi_2=\arcsin j$ (up to an arbitrary multiple of $2\pi$).

For sufficiently large $j$, the quantum degree of freedom $\varphi_2$
undergoes tunneling to a new local minimum nearby $2\pi n$, while the
classical degree of freedom $\varphi_1 =\arcsin j$ remains localized, thus
allowing a flux $\simeq n\Phi_0,\ n \in \mathbb{N}$ to enter the SQUID loop,
cf.  Fig.\ \ref{fig:effpot}. If the resulting force on the classical
phase $\varphi_1$ is sufficiently large, the M\"unchhausen decay is enabled
with a classical decay of $\varphi_1$ and successive iteration of quantum
decay (directed along $\varphi_2$, flux entry) and classical relaxation
(directed mainly along $\varphi_1$, flux exit), cf. Fig.\ \ref{fig:corrpot}.
\begin{figure}[htb]
\includegraphics{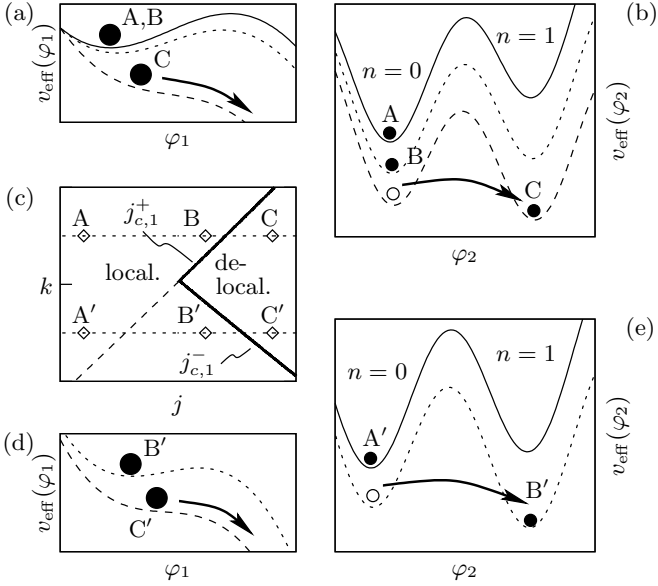}%
\centering
\caption{Illustration of the assembly of $j_c(k)$ from the segments
$j_{c,1}^+(k)$, Eq.\ \eqref{eq:firstmin}, and $j_{c,1}^-(k)$, Eq.\
\eqref{eq:firstforce}. In (c), the relevant region of the $j$-$k$-plane is
shown. At A, the quantum junction (see (b)) has relaxed to the initial minimum
($n=0$) and the classical junction is localized in a stable minimum (see (a)).
Increasing the bias current $j$ and approaching B, the potential barrier
confining the classical phase in $v_ \mathrm{eff}(\varphi_1)$ is only slightly
reduced. As $j>j_{c,1}^+(k)$ (C), the increase in $j$ has deformed the
effective potential  $v_\mathrm{eff}(\varphi_2)$ as to allow a phase slip of
$\varphi_2$ (see (b)). The additional force it exerts on the classical
junction immediately removes the barrier (see (a)) and delocalizes $
\varphi_1$ at $j_c(k)=j_{c,1}^+(k)$. On the other hand, an increase in bias
$j>j_{c,1}^+(k)$ from A$'$ to B$'$ at lower $k$ leads to a phase slip of
$\varphi_2$ (see (e)), without delocalizing $\varphi_1$ (see (d)). A further
increase in $j$ triggers a classical decay (see (d)) when crossing the
critical line at $j_c(k)=j_{c,1}^-(k)$ (C$'$).}
\label{fig:explanation} 
\end{figure}

The strong coupling at large $k$ keeps the local minimum at $\varphi_2=\arcsin
j$ lower in energy than the adjacent local well at $\varphi_2\approx 2 \pi$
for all bias currents $j$, hence tunneling is inhibited and no
M\"unchhausen decay takes place.  At lower $k$, a bias current $j<1$ can
sufficiently lower the adjacent well such as to bring both minima to equal
height. This condition is reached once the minimum of the parabola in
$v_\mathrm{eff}(\varphi_2)$ (at $\varphi_2=\arcsin j+j/k$) is aligned with the
midpoint between the two corresponding minima of $-\cos \varphi_2$ (at
$\varphi_2=\pi$); i.e., if $\arcsin j+j/ k=\pi$. Thus, for $k<k_{c,1}^+(j)$
with\cite{hallatschek:2000}
\begin{equation}
   k_{c,1}^+(j)=\frac{j}{\pi-\arcsin j},
   \label{eq:firstmin}
\end{equation}
the minimum at $\varphi_2\approx 2\pi$ is lower and a quantum decay is enabled
(cf.\ Fig.\ \ref{fig:explanation}(b), point C in the diagram). The jump of
$\varphi_2$ by roughly $2\pi$ then pulls the heavy junction out of its minimum
and the M\"unchhausen decay is initiated, cf.\ Figs.\ \ref{fig:explanation}(a) and
(b) associated with the sequence of points A, B, and C in Fig.\
\ref{fig:explanation}(c). The smaller $k$ becomes, i.e., the weaker
$\varphi_2$ is bound to $\varphi_1$, the less current is needed to lower the
adjacent minimum such as to allow for a decay of $\varphi_2$, hence the
positive slope of $k_{c,1}^+(j)$.

Decreasing $k$ too far, however, the pulling force exerted by the quantum
junction may not be sufficient to drag the heavy junction out of its minimum.
Hence, we have to investigate the shape of the potential $v(\varphi_1,
\varphi_2)$ after the phase slip in $\varphi_2$, i.e., nearby the point
$\varphi_1\gtrsim \arcsin j,\ \varphi_2\approx 2 \pi$, and check whether the
barrier against a classical decay (mainly along $\varphi_1$) has disappeared;
this is identical to the calculation of the critical current of a SQUID
with a trapped flux.\cite{Tsang:1975, Lefevre-Seguin:1992}

\begin{figure}[htb]
\includegraphics{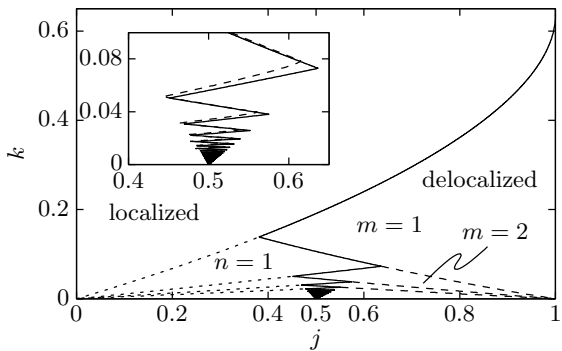}%
\centering
\caption{Phase diagram of the dynamically asymmetric dc-SQUID as a function of
bias current $j=J/2 J_c$ and inductive coupling $k=\Phi_0 c/2\pi  LJ_c$. Here,
we assume strong damping $\alpha_{p,1}\gg1$ and $\alpha_{p,2}>1$. The
effective critical current $j_c(k)$ (solid line) marks the boundary between a
localized classical junction (lower bias $j < j_c$) and a delocalized
classical junction ($j > j_c$), corresponding to a finite voltage state of the
SQUID. Branches with negative slope are determined by a classical instability
(mainly along $ \varphi_1$), while those with positive slope are determined by
a quantum instability of the light junction.  For $j<j_c(k)$, the dotted lines
$j_{c,n}^+(k)$ mark the entry of flux through the quantum junction (the
integer $n$ approximately quantifies the flux through the ring in
the stable state); these lines can be measured via monitoring of the flux
threading the loop. For $j>j_c(k)$, the dashed lines $j_{c,m}^-(k)$ mark the
minimum number $m$ of flux units necessary to delocalize the classical
junction. The inset shows a comparison between the approximate result, Eqs.\
\eqref{eq:globmin} and \eqref{eq:approximate_critical_k} (solid line) and the
exact numerical result (dashed).}
\label{fig:phase_diagram_vv}
\end{figure}

We first determine the position of the minimum by solving the equations
\begin{eqnarray}
   \partial_{\varphi_1}v(\varphi_1,\varphi_2)&=&\sin(\varphi_1)-j+k(\varphi_1
   -\varphi_2)=0,
   \label{eq:min_1}\\
  \partial_{\varphi_2}v(\varphi_1,\varphi_2)&=&\sin(\varphi_2)
   -j-k(\varphi_1
   -\varphi_2)=0,
   \label{eq:min_2}
 \end{eqnarray}
for $\varphi_1\gtrsim \arcsin j$ and $\varphi_2\approx 2 \pi$. At the critical
coupling $k_{c,1}^-(j)$ the minimum should merge with a saddle and define an
inflection point along some direction in the $(\varphi_1,\varphi_2)$-plane.
The resulting system of equations requires numerical solution and the result
is shown in the inset of Fig.\ \ref{fig:phase_diagram_vv}. However, within the
interesting region at small coupling $k$ we can find an approximate analytical
solution: For $k\ll 1$, the side minimum of $v(\varphi_1,\varphi_2)$ becomes
unstable predominantly along the $\varphi_1$-direction. Thus, the minimum
disappears if
\begin{equation}
\partial_{\varphi_1}^2v(\varphi_1,\varphi_2)=\cos\varphi_1+k=0.
\label{eq:phi1}
\end{equation}
We choose the solution $\varphi_1\approx\pi/2$ (we set $k=0$ and assume $0\leq
\varphi_1< 2 \pi$; the other solution $\varphi_1\approx 3\pi/2$ cannot solve
Eq.\ \eqref{eq:min_1} and is discarded). Inserting $\varphi_1$ into the sum of
Eqs.\ \eqref{eq:min_1} and \eqref{eq:min_2} yields the relation
\begin{equation}
\sin\varphi_2\approx2j-1,
\end{equation}
from which we find $\varphi_2\approx2\pi+\arcsin[2j-1]$ (the other solution
$\varphi_2\approx3\pi-\arcsin[2j-1]$ is excluded since it describes 
a maximum along the $\varphi_2$-direction).  Inserting $\varphi_1$ and
$\varphi_2$ into Eq.\ \eqref{eq:min_1}, we find the condition
\begin{equation}
   k_{c,1}^-(j)\approx\frac{1-j}{(3/2)\pi+\arcsin(2j-1)}.
   \label{eq:firstforce}
\end{equation}
For $k>k_{c,1}^-(j)$ no barrier blocks the motion of $\varphi_1$ after the
phase slip in $\varphi_2$ and a classical decay of $\varphi_1$ is enabled,
thus completing the M\"unchhausen decay. This scenario is described by the
sequence A, B, C in Fig.\ \ref{fig:explanation}(c). For $k < k_{c,1}^-(j)$ the
force after the phase slips is too small and an additional increase in
$j$ is necessary to drive the system overcritical, as illustrated by the
sequence A$'$, B$'$, C$'$ in Fig.\ \ref{fig:explanation}(c) and Figs.
\ref{fig:explanation}(d) and (e). The increase in critical current with
decreasing coupling defines a negative slope for $k_{c,1}^-(j)$.

We define the effective critical current $j_c(k)$ as the phase boundary
between the stable region ($j<j_c(k)$), where the M\"unchhausen decay is
prohibited, and the delocalized phase ($j>j_c(k)$). This critical line is
assembled from the segments $j_{c,1}^\pm(k)$, the inverse functions of
$k_{c,1}^\pm(j)$, Eqs.\ \eqref{eq:firstmin} and \eqref{eq:firstforce},
respectively: Ramping up the current $j$ at large values of $k$, we eventually
cross $j_{c, 1}^+(k)$. The phase slip of $\varphi_2$ immediately enables the
classical decay of $\varphi_1$ and the system enters a running state, hence,
$j_c(k)=j_{c,1}^+(k)$. At lower $k$, increasing $j$ beyond $j_{c,1}^+(k)$
triggers a phase slip of $\varphi_2$, but the resulting force is too weak to
delocalize $\varphi_1$.  A further increase in $j$ is necessary until, at
$j=j_{c,1}^-(k)$, the minimum disappears and the M\"unchhausen decay proceeds,
hence, $j_c(k)=j_{c,1}^-(k)$.

Following the above discussion, the system is stable for $j<j_{c,1}^-(k)$.
Upon further decreasing $k$, however, more and more side minima in
$v_\mathrm{eff} (\varphi_2)$ become accessible to the quantum junction. From
Fig.\ \ref{fig:effpot} we notice, that the different local minima of
$v_\mathrm{eff} (\varphi_2)$ are located near $\varphi_2\approx 2\pi n$. They
describe the state where a flux $\simeq n\Phi_0$ has entered the loop and we
label them by the index $n$.  Eq.\ \eqref{eq:firstmin} is then
straightforwardly generalized to the critical line $k_{c,n}^+(j)$ describing
the entry of the $n$-th fluxon. The $n$-th phase slip of $\varphi_2$ occurs
when the global minimum in $v_\mathrm{eff}(\varphi_2)$ shifts from
$\varphi_2\approx 2\pi(n-1)$ to $\varphi_2\approx 2\pi n$. We can proceed
analogously to the derivation of Eq.\ \eqref{eq:firstmin}: Let
$\varphi_1^{\mathrm{min},n-1}$ be the solution of Eqs.\ \eqref{eq:min_1} and
\eqref{eq:min_2} near $\varphi_1\gtrsim\arcsin j,\ \varphi_2\approx 2\pi
(n-1)$.  
In order to find the crossing in the height of the two minima we account for 
the relaxation of $\varphi_1$ and align the minimum of the shifted parabola 
($\varphi_2=\varphi_1^{\mathrm{min},n-1}+j/k$) with the midpoint between the 
two minima of the $-\cos\varphi_2$-potential ($\varphi_2=(2n-1)\pi$), hence
$\varphi_1^{\mathrm{min},n-1}+j/k=(2n-1)\pi$ and
\begin{equation}
   k_{c,n}^+(j)= \frac{j}{(2n-1)\pi-\varphi_1^{\mathrm{min},n-1}}.
   \label{eq:exact_critical_i}
\end{equation}
A convenient approximation is made by ignoring the relaxation of $\varphi_1$, 
resulting in the expression
\begin{equation}
   k_{c,n}^+(j)\approx\frac{j}{(2n-1)\pi-\arcsin j}.
   \label{eq:globmin}
\end{equation}

Similarly, we can generalize Eq.\ \eqref{eq:firstforce} to a critical
coupling $k_{c,n}^-(j)$, determining whether the resulting force at given
fluxon index $n$ is sufficient to delocalize $\varphi_1$. We then solve
Eqs.\ \eqref{eq:min_1} and \eqref{eq:min_2} near $\varphi_1\approx \arcsin j,\
\varphi_2\approx 2\pi n$, proceed as in the derivation of Eq.\
\eqref{eq:firstforce}, and find (for small $k\ll 1$)
\begin{equation}
   k_{c,n}^-(j)\approx\frac{1-j}{(2n-1/2)\pi+\arcsin(2j-1)}.
   \label{eq:approximate_critical_k}
\end{equation}

The critical current line $j_c(k)$ is constructed from interchanging segments
of $j_{c,n}^+(k)$ and $j_{c,n}^-(k)$, resulting in the dynamical phase
diagram, Fig.\ \ref{fig:phase_diagram_vv}. Note that the different nature of
the decay, classical or quantum, associated with the two types of critical
lines may allow for an experimental distinction: Ramping the current past a
$^+$-type segment of $j_c(k)$ triggers a \emph{quantum} decay with a broad
histogram describing multiple measurements. A $^-$-type segment of $j_c(k)$
triggers a \emph{classical} decay with a sharp histogram (the quantum decay of
$\varphi_2$ needs to have occurred already, which limits the ramping speed
before reaching the critical line). Note that the lines $j_{c,n}^+(k)$ are
detectable throughout all the stable portion of the phase diagram, e.g., via a
measurement of the flux threading the loop (the flux increases by
approximately one flux unit upon crossing the dotted lines in Fig.\
\ref{fig:phase_diagram_vv}).

The phase diagram in Fig.\ \ref{fig:phase_diagram_vv} shows that the critical
line $j_c(k)$ approaches the value $1/2$ for $k\rightarrow 0$. This is understood 
from analyzing Eqs.\ \eqref{eq:globmin} and
\eqref{eq:approximate_critical_k} for large $n$ and small $k$, providing the
relations $k_{c,n}^+(j)\approx j/2 n \pi$ and $k_{c,n}^-(j)\approx (1-j)/ 2 n
\pi$, respectively. Equating the two conditions gives $j_c\approx 0.5$. From a
physical point of view, this result can be easily explained: The decay of
$\varphi_2$ proceeds towards the bottom of the parabola in
$v_\mathrm{eff}(\varphi_2)$; for $k\rightarrow 0$, the current through the
quantum junction then approaches zero, $J_2\propto\sin\varphi_2\approx 0$.
Consequently, all current is redirected through the classical junction 1,
with the effective bias now increased to $J_1=2 jJ_c$. Junction 1 thus
turns dissipative at $j=1/2$, the critical current of a single junction but
half the critical current of the dc-SQUID.
\begin{figure}[htb]
\includegraphics{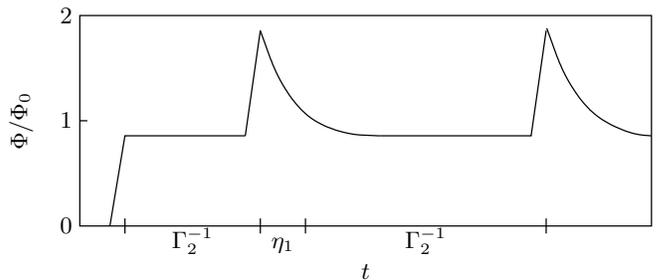}%
\centering
\caption{Sketch of a time-trace of the flux threading the SQUID loop during
the decay process of $\varphi_1$ and $\varphi_2$ at $j > j_c(k)$, cf.\ Fig.\
\ref{fig:corrpot}(b). The flux increases by approximately one flux unit each
time the quantum junction decays; this process then involves the tunneling
rate $\Gamma_2$ of $\varphi_2$, Eq.\ \eqref{eq:gamma}.  The additional flux
leaves the loop again within the time interval $\sim \eta_1$ as the classical
junction relaxes dissipatively to the next minimum.}
\label{fig:flux}
\end{figure}

The dynamical phase diagram, Fig.\ \ref{fig:phase_diagram_vv}, can be tested
experimentally through a measurement of the dc-voltage drop across the device.
The second Josephson relation\cite{Josephson:1962} tells that the
time-averaged voltage $\langle V\rangle_t\propto \langle \dot\varphi_1
\rangle_t= \langle\dot\varphi_2\rangle_t$. Inspecting 
Fig.\ \ref{fig:corrpot}(b), we understand that the continuous iteration of quantum 
and classical
decays of the light and heavy phase variables generate the nonzero averages
$\langle \dot\varphi_1\rangle_t =\langle \dot\varphi_2 \rangle_t$ for
$j>j_c(k)$. For the overdamped setup as described above, the voltage drop is
small as strong damping reduces the tunneling rate determining
$\langle\dot\varphi_2\rangle_t$. This is particularly true close to the
critical line, where the variable $\varphi_2$ needs to tunnel in the `flat'
part of $v_\mathrm{eff}(\varphi_2)$. The decay process is faster for weak
damping and thus generates a larger voltage signal---we discuss this
situation in the next Section. 

Another characteristic signal of the M\"unchhausen decay is the time-trace of
the magnetic flux threading the SQUID loop during the alternating decay of the
two phases (cf.\ Fig.\ \ref{fig:corrpot}(b)), see Fig.\ \ref{fig:flux} for an
illustration.  Such time-traces exhibit two characteristic time scales, one
due to the tunneling of the quantum junction involving the rate $\Gamma_2$
separating subsequent peaks of flux entry into the loop; the other is due to the
classical relaxation of the heavy junction and involves the dissipative time
$\sim\eta_1$ describing the flux exit and return to the low-flux state of
the loop. At small $k$ and small effective bias between
two minima,
\begin{equation}
\frac{\Gamma_2^{(n)}}{\omega_{0,2}}\approx \frac{\hbar
\omega_{0,2}E_{c,2}}{E_J^2}\left(\frac{R_Q}{\pi R_2}\right)^{7/2} 
(j-j_{c,n}^+)^{2 R_Q/R_2-1}, \label{eq:gamma}
\end{equation}
with $R_Q=h/4e^2$.\cite{korshunov:1987} 

\section{Weak damping}
\label{sec:underdamped}

\begin{figure}[htb]
\begin{center}
\includegraphics{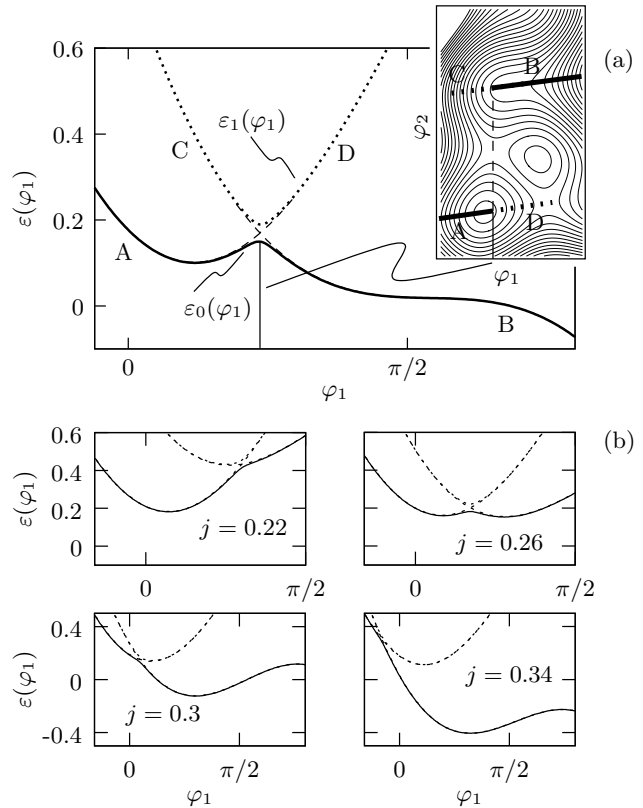}
\caption{Illustration of the effective potential for the `heavy' phase
$\varphi_1$, generated by the two lowest levels of the `light' phase. In (a),
the situation for $k=0.15,\ j=0.36$ is shown. The avoided level crossing
$2\Delta_2$ in the spectrum of the light junction is reflected in the effective
potentials $\varepsilon_0(\varphi_1)$ (solid) and $\varepsilon_1(\varphi_1)$
(dotted) for the heavy junction.  The inset illustrates the 2-dimensional path
$[\varphi_1,\varphi_2^{\mathrm{min},0/1}(\varphi_1)]$ in $v(\varphi_1,
\varphi_2)$ corresponding to $\varepsilon_0(\varphi_1)$ (solid, A/B) and
$\varepsilon_1(\varphi_1)$ (dotted, C/D); the upward/downward jump along
$\varphi_2$ to the next minimum corresponds to the avoided crossing in
$\varepsilon(\varphi_1)$. (b) Series of $\varepsilon_0(\varphi_1)$ (solid) and
$\varepsilon_1(\varphi_1)$ (dotted) for various values of $j$ at $k=0.1$.
Increasing $j$ shifts the position of the avoided crossing in $\varepsilon
(\varphi_1)$ to the left. Its transition past $\varphi_1^\mathrm{min}$
(between $j=0.26$ and $j=0.3$) corresponds to crossing the line $j_{c,1}^+(k)$
and implies the decay of the quantum junction to the next local minimum.
Note that in the present case the newly formed minimum is stable (a point of
type B$'$ in Fig.\ \ref{fig:explanation}).}
\label{fig:p1}
\end{center}
\end{figure}

The behavior of the system for weak damping $\alpha_{LC,2}\ll 1,\
\alpha_{p,i}\ll 1$ is quite similar to the one encountered before, while the
(small) differences strongly depend on the many system parameters and their
associated timescales. To begin with we neglect dissipative effects.
Bearing in mind the limit of large $C_1/C_2$, we can exploit the adiabatic
separation of fast and slow degrees of freedom; i.e., we first consider the
fast problem for the quantum junction,
\begin{equation}
\mathcal{H}_2\Psi_{l}(\varphi_2)=\varepsilon_l \Psi_l(\varphi_2),
\label{eq:ev}
\end{equation}
with the reduced Hamiltonian (cf.\ Eq.\ \eqref{eq:motion})
\begin{equation}
  \mathcal{H}_2=-4E_{c,2}\partial_{\varphi_{2}}^2
   + E_{J} v_\mathrm{eff}[\varphi_{1}] (\varphi_{2})
\end{equation}
defined at \emph{fixed} $\varphi_1$. Eq.\ \eqref{eq:ev} establishes the
dynamics of the `light' variable $\varphi_2$ and we can find the energies
$\varepsilon_l(\varphi_1)$. These act as effective potentials for the `heavy'
degree of freedom $\varphi_1$; e.g., assuming the quantum junction to reside
in a state $|l\rangle$, $\langle \varphi_2| l \rangle = \Psi_l(\varphi_2)$,
the effective potential for $\varphi_1$ is given by $\varepsilon_l
(\varphi_1)$. Typical examples for the effective potentials $\varepsilon_0
(\varphi_1)$ (ground state) and $\varepsilon_1(\varphi_1)$ (first excited
state) are shown in Fig.\ \ref{fig:p1}.

In the dissipative situation, we could derive the phase diagram from simply
analyzing the potential $v(\varphi_1,\varphi_2)$; here, instead, we first
determine which states $l$ of the light junction are relevant and then analyze
the dynamics of $\varphi_1$ in the resulting effective potential (e.g., Fig.\
\ref{fig:p1}(a)) defined via $\varepsilon_l(\varphi_1)$, thus 
determining whether the
heavy junction stays localized or enters a running state. Both tasks strongly
depend on the various timescales involved.  We start with the timescale of the
`external control', the rate at which the bias $j$ is ramped. Whereas the case
of strong damping excluded the presence of kinetic energy in the system, this
is no longer the case here. The importance of the ramping rate then comes
about through the amount of energy which is transferred to the two degrees of
freedom. The change of $j$
translates to a change of the potential $v(\varphi_1,\varphi_2)$ with a rate
of the order of $\partial_t j$. The ramping can be adiabatic, in which case no
energy is given to the system, or instantaneous, where the amount of
transferred energy is maximal. Intermediate types of ramping will not be
discussed.

The adiabatic separation of time scales requires a large capacitance ratio
$C_1\gg C_2$.  For $\omega_{p,1} \ll \omega_{p,2}$ the intravalley motion of
the two junctions are separated. The (tunneling) motion of the light junction
across valleys produces a stronger condition: Consider 
the tunneling splitting $\Delta_2$ in the spectrum of
$\mathcal{H}_2$ at an avoided level crossing, cf.\ Fig.\ \ref{fig:p1}(a),
\begin{equation}
   \Delta_2\sim\hbar\omega_{p,2}
   \exp \bigl(-\gamma\sqrt{{E_J}/{4E_{c,2}}}\bigr),
\label{eq:delta}
\end{equation}
with $\gamma$ depending on the actual shape of the potential,
\begin{equation}
   \gamma=\int_a^b d\varphi_2 \sqrt{v_\mathrm{eff}(\varphi_2)-E_0/E_J}
\end{equation}
in the quasi-classical approximation. Here, $E_0$ is the ground state energy
in the well and $a$ and $b$ are the left and right classical turning points,
respectively, $v_\mathrm{eff}(\varphi_2=a) = v_\mathrm{eff}(b) = E_0/E_J$.
The tunneling gap $\Delta_2$ then is relevant when analyzing the motion of the
classical junction ($\varphi_1$) in the effective potential given by
$\varepsilon_l(\varphi_1)$. If the motion of $\varphi_1$ is sufficiently slow,
its passage along an avoided crossing in the spectrum of $\mathcal{H}_2$ is
adiabatic and $\varphi_1$ follows $\varepsilon_l(\varphi_1)$ through the
anticrossing (cf.\ the trajectory A$\,\to\,$B in Fig.\ \ref{fig:p1}(a)); i.e.,
the phase $\varphi_2$ tunnels to the next adjacent minimum. On the other hand,
if the motion of $\varphi_1$ is fast, the state of the quantum junction
undergoes Landau-Zener tunneling (cf.\ the trajectory A$\,\to\,$D in Fig.\
\ref{fig:p1}(a); the light junction then remains trapped in its local well)
and $\varphi_1$ follows the effective potential $\varepsilon_{l\pm1}
(\varphi_1)$ after the avoided crossing. The probability $p_\mathrm{LZ}$ for
Landau-Zener tunneling at an anticrossing of the two lowest levels of
$\mathcal{H}_2$ is given by
\begin{equation}
  p_\mathrm{LZ}=\exp\left(-\frac{2\pi \Delta_2^2}
  {\hbar |d_t \varepsilon_{10}(t)|} \right),
\label{eq:lz}
\end{equation}
where $\varepsilon_{10}(t)=\varepsilon_1(t) - \varepsilon_0(t)$ and $d_t$ is
the total time derivative. To estimate the rate of change in energy
$d_t\varepsilon_{10}(t)$, we start from
\begin{equation}
\varepsilon_{10}\approx E_J[v(\varphi_1,\varphi_2^{\mathrm{min},n+1})-
v(\varphi_1,\varphi_2^{\mathrm{min},n})]
\end{equation}
and find the time derivative 
\begin{eqnarray}
d_t\varepsilon_{10}(t) &=&\partial_{\varphi_1}\varepsilon_{10} \dot\varphi_1
\nonumber\\
&&+E_J\big[\partial_{\varphi_2}v(\varphi_1,\varphi_2)|_{\varphi_2^{\mathrm{min},n+1}}
\partial_{\varphi_1}\varphi_2^{\mathrm{min},n+1}\nonumber\\
&&-\partial_{\varphi_2}v(\varphi_1,\varphi_2)|_{\varphi_2^{\mathrm{min},n}}
\partial_{\varphi_1}\varphi_2^{\mathrm{min},n}
\big]\dot\varphi_1 
\nonumber\\
&=&kE_J (\varphi_2^{\mathrm{min},n}-\varphi_2^{\mathrm{min},n+1})\dot
\varphi_1,
\end{eqnarray}
where we have used that $\partial_{\varphi_2}v(\varphi_1,\varphi_2)$ vanishes
at the minima $\varphi_2^{\mathrm{min},n}$ and only the term originating from
the coupling term in $v(\varphi_1,\varphi_2)$ is relevant. Since
$\dot\varphi_1$ is at most of order $\omega_{0,1}$, we obtain the estimate $d_t
\varepsilon_{10}\sim2\pi k \omega_{0,1} E_J$. The smallness of 
$p_{LZ}$, guaranteeing adiabatic motion of 
the classical junction (this requires a
correspondingly large $C_1$) then follows from
\begin{equation}
   k \frac{E_J}{ \Delta_2}\frac{\hbar\omega_{0,1}}{ \Delta_2} \sim k
   \sqrt{\frac{E_{c_1}}{E_{c_2}}}\sqrt{\frac{E_J}{E_{c_2}}} \exp\Bigl(2\gamma
   \sqrt{\frac{E_J}{4E_{c_2}}} \Bigr) \ll 1.
\label{eq:landau} 
\end{equation}

\begin{figure}[htb]
\begin{center}
\includegraphics{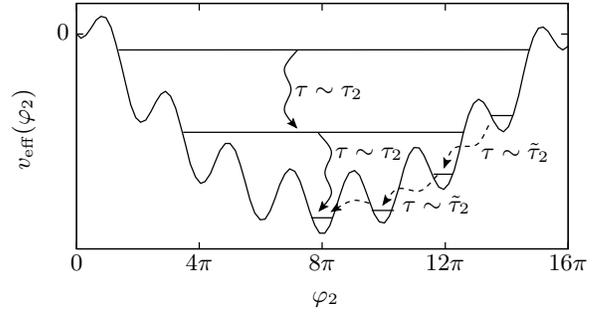}
\caption{Illustration of decay sequences of an excited `superwell'-state of
the quantum junction in the effective potential $v_\mathrm{eff}(\varphi_2)$,
Eq.\ \eqref{eq:effpot}. Solid arrows indicate a decay sequence involving only
states within the `superwell' with a typical decay time $\tau_2$. The dashed
arrows illustrate another (extreme) decay sequence, involving localized states
in the side-minima of $v_ \mathrm{eff}(\varphi_2)$. The decay time then
involves the typical lifetime $\tilde \tau_2 \gg \tau_2$ of local ground
states within side-wells. }
\label{fig:decay}
\end{center}
\end{figure}

The relaxation of the quantum degree of freedom $\varphi_2$ is another
important element in our discussion. The dissipation described by the normal
resistance $R$ in the RCSJ-equation of motion, cf.\ Eq.\ \eqref{eq:motion},
leads to typical finite lifetimes $\tau_2 \sim RC_2= 1/2\alpha_2
\omega_{p,2}$ of the excited states $|l \rangle$ of the quantum
junction.\cite{esteve:1986}  A much longer lifetime $\tilde\tau_2$ shows up if
the quantum junction is trapped in a local ground state of a side-well in
$v_\mathrm{eff}(\varphi_2)$, cf.\ Fig.\ \ref{fig:decay}. The decay then is
protected through a large barrier $E_b$, enhancing the typical lifetime to a
value $\tilde \tau_2$, with $\tau_2 \ll \tilde \tau_2 \propto
 \exp(2\gamma\sqrt{E_J/4E_{c,2}})$.\cite{Averin:2000}
Thus, there are two (extreme) ways how a highly exited state in the
`superwell' of $v_\mathrm{eff}(\varphi_2)$ can decay, see Fig.\
\ref{fig:decay}, either via states within the superwell involving the typical
lifetime $\tau_2$, or via states involving a tunneling process, resulting in
an exponentially larger decay time of the order of $\tilde\tau_2$.

In the following, we investigate three regimes and derive the corresponding
effective critical current $j_c(k)$. We begin with the case of adiabatic
ramping (Sec.\ \ref{sec:adiabatic}), where we assume that the ramping is
slower than the typical relaxation time of the heavy junction and the
timescales between $\varphi_1$ and $\varphi_2$ separate; i.e., inequality 
\eqref{eq:landau} holds. We show that the
critical current $j_c(k)$ obtained in the dissipative case, Sec.\
\ref{sec:overdamped}, is only slightly altered due to the (small) finite
amount of kinetic energy the classical junction can store. Next, we discuss
the case of fast ramping (Sec.\ \ref{sec:fast}), where the phase $\varphi_1$
remains effectively `frozen' during the current ramping and quite an
appreciable amount of potential energy is converted to kinetic energy 
in the motion of the classical junction,
leading to a pronounced reduction of $j_c(k)$. Both discussions require a very
large value of $C_1/C_2\rightarrow \infty$ to guarantee the absence of Landau-Zener 
tunneling due to the motion of $\varphi_1$. 
Subsequently, we discuss the consequences of a moderate ratio
$C_1/C_2\lesssim 10^4$ accessible in today's experiments in the last part of
this section (Sec.\ \ref{sec:farthest}). Then, $\varphi_2$ can remain trapped 
in a localized state in a side well during the motion and liberation of the heavy 
junction. The probabilistic effects emerging
for this situation change the nature of the phase diagram qualitatively.

\subsection{Adiabatic ramping of the bias current}
\label{sec:adiabatic}

We start from the system with fixed coupling $k$ at $j=0$ and consider a state
localized at $\varphi_1=\varphi_2=0$. We assume slow current ramping on the
dissipative timescale of the classical junction's motion,
\begin{equation}
\partial_t j \ll 1/\tau_1,
\label{eq:adiabatic}
\end{equation}
where $\tau_1\sim RC_1$ denotes the classical relaxation time of the heavy
junction.  The inequalities \eqref{eq:adiabatic} (slow ramping) and
\eqref{eq:landau} (slow motion of $\varphi_1$) guarantee that the quantum
junction remains in the global ground state, while the classical junction
follows its local ground state near $\varphi_1 = \arcsin j$.

The effective potential for the heavy junction then is determined by the
ground state energy $\varepsilon_0 (\varphi_1)$. In our estimate
\begin{equation}
   \varepsilon_{0}(\varphi_{1})\approx E_{J}v[\varphi_{1},\varphi_{2}^\mathrm{glob}
   (\varphi_1)]
\label{eq:e_0}
\end{equation}
we neglect the correction due to the ground state energy
$E_{0}\approx\hbar\omega_{p,2}/2$ (as measured from the bottom of the
potential) of the light junction;\cite{footnote3}
the phase $\varphi_{2}^\mathrm{glob}
(\varphi_1)$ refers to the global minimum of the light junction. Furthermore,
we ignore the small splitting $2\Delta_2$ at the avoided level crossing (cf.\
Fig.\ \ref{fig:p1}(a)), replacing it with a sharp kink.

With the above approximations, the force acting on the heavy junction
originates from the classical force exerted by the quantum junction residing
in the global minimum.  The situation is thus the same as in the strong
damping regime: the `flux-entry lines' $j_{c,n}^+(k)$, marking the current
where the light junction is allowed to tunnel, are still given by Eq.\
\eqref{eq:exact_critical_i}.  When the classical junction has relaxed to the
bottom of a minimum, that minimum has to disappear in order for $\varphi_1$ to
enter the running state. Given the expression \eqref{eq:e_0} for
$\varepsilon_0(\varphi_1)$, the local minimum in $\varepsilon_0(\varphi_1)$
corresponding to the $n$-th side minimum in $v_\mathrm{eff} (\varphi_1,
\varphi_2)$ turns into an inflection point at $j=j_{c,n}^-(k)$, Eq.\
\eqref{eq:approximate_critical_k}. This condition again agrees with the
one for strong damping.
\begin{figure}[htb]
\begin{center}
\includegraphics{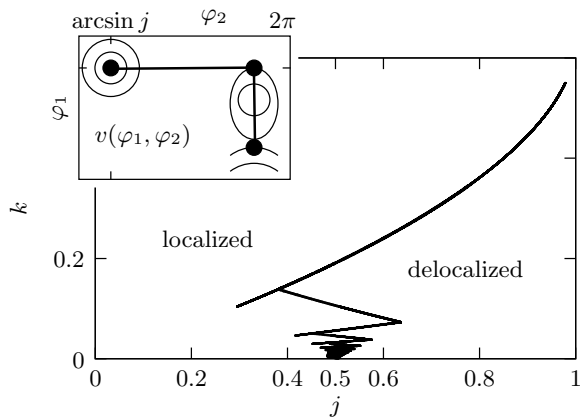}
\caption{Phase diagram of the dynamically asymmetric dc- SQUID for weak
damping and adiabatic ramping, $(\partial_t  j \ll
\tau_1^{-1}$). The critical line $j_c(k)$ is nearly identical with the one in
the strong damping case, but is hysteretic along extensions of the critical
lines $j_{c,n}^+(k)$. These are due to the nonzero kinetic energy acquired by
the classical junction when sliding down in the new well formed after the
tunneling of $\varphi_2$ at $j_{c,n}^+(k)$ and allowing $\varphi_1$ to
surmount a residual potential barrier in the effective potential
$\varepsilon_0(\varphi_1)$. The inset illustrates the condition determining
the termination point of the hysteretic line along $j_{c,1}^+(k)$. The three
solid dots mark points of equal potential energies; the energy gained by the
classical junction due to the phase slip of $\varphi_2$ just suffices to climb
the top of the remaining barrier along $\varphi_1$.}
\label{fig:phase_diagram_adiabatic}
\end{center}
\end{figure}

A difference to the previous dissipative situation is given by the possibility
to transform potential into kinetic energy. Ramping $j$ at fixed $k$ past the
flux-entry line $j_{c,n}^+(k)$ triggers a phase slip in $\varphi_2$ and
deforms the potential well in $\varepsilon_0(\varphi_1)$ for the classical
junction. The latter finds itself on the slope of the well and gains kinetic
energy while sliding down towards the new minimum.  The classical junction
then may enter a running state if this kinetic energy gain is sufficient to
surpass the barrier blocking the newly formed ($n$-th) minimum.  Different
from the overdamped case, the system then exhibits hysteretic behavior, as
illustrated in the dynamical phase diagram Fig.\
\ref{fig:phase_diagram_adiabatic}, where the running state can be entered
along extensions of the $j_{c,n}^+(k)$ line away from the phase boundary.  The
point where these extensions terminate is determined by the condition that the
energy right after the tunneling of $\varphi_2$ is equal to the energy at the
top of the barrier, as illustrated in the inset of Fig.\
\ref{fig:phase_diagram_adiabatic}. If the coupling $k$ is too small to lower
the barrier sufficiently (cf.\ Fig.\ \ref{fig:p1} (b)), the state of the heavy
junction upon reaching $j=j_{c,n}^+(k)$ is transformed into a localized
excited state of the newly formed potential well in $\varepsilon_0
(\varphi_1)$.  Given the slow ramping $\partial_t j\ll
\tau_1^{-1}$, the heavy junction relaxes to the ground state of the well and a
further increase in $j$ is necessary to remove the barrier completely, hence
the critical line jumps forward to $j_{c}(k)=j_{c,n}^-(k)$.

\subsection{Fast ramping of the bias current}
\label{sec:fast}

We now turn to the case $\partial_t j \gg \omega_{p,1}$, where the ramping of
$j$ is instantaneous with respect to the (massive) dynamics of the heavy
junction; the final current $j$ has to remain below $\sim 0.7$, since fast
ramping beyond this value allows the classical junction to overcome the
potential barriers through conversion of potential to kinetic energy.
Furthermore, we study the setting where $\omega_{p,1}\ll \tau_2^{-1},
\tilde\tau_2^{-1}$; the quantum junction then has relaxed to the ground state
(at $\varphi_2^{\mathrm{min},n}$) in the effective potential $v_
\mathrm{eff}(\varphi_2)$ before any motion of the heavy junction sets in.
Hence, initially, the effective potential for the classical junction is given
by the ground state energy $\varepsilon_0(\varphi_1)$.  The absence of Landau-Zener
tunneling (due to the inequality \eqref{eq:landau}) then assures that the
effective potential is given by $\varepsilon_0(\varphi_1)$ during the entire
motion of the classical junction.

\begin{figure}[htb]
\begin{center}
\includegraphics{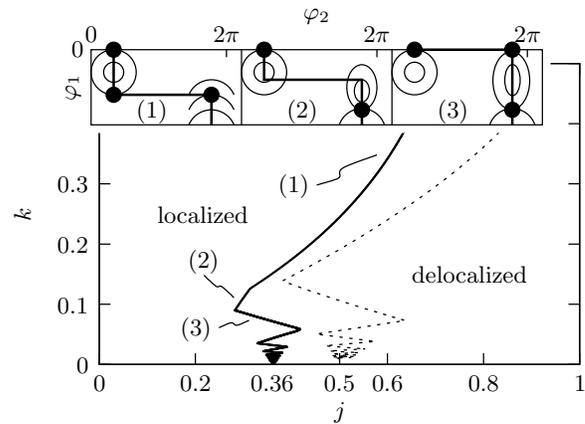}
\caption{Effective critical current $j_c(k)$ (solid) of the current-biased
dc-SQUID with strongly asymmetric parameters as a function of bias current $j$
and inverse inductance $k\propto 1/L$. We assume instantaneous ramping
of the bias $j $ and a strictly classical and adiabatically slow (as compared
to the dynamics of the quantum junction) motion of the weakly damped heavy degree
of freedom $ \varphi_1$.  For comparison, the critical line of the
dissipative case is also displayed (dashed line). The inset illustrates the
three segments constituting the first part of $j_c(k)$, where solid dots
indicate points of equal potential energy.  Inset (3) shows the
situation right at the tip between segments (2) and (3).}
\label{fig:mmfa}
\end{center}
\end{figure}

We then can find the criterion for $\varphi_1$ to enter a running state:
During the fast ramping of the current $j$, the heavy junction remains frozen
at $\varphi_1=0$. We have to check whether the potential energy of the heavy
junction is sufficient to overcome all barriers in
$\varepsilon_0(\varphi_1>0)$; i.e., we have to inspect if the maximum of
$\varepsilon_0(\varphi_1)$ in the interval $0< \varphi_1<2 \pi$ is realized at
$\varphi_1=0$. As soon as this condition applies, the classical junction
becomes delocalized. This analysis has to be performed numerically, and the
result is displayed in Fig.\ \ref{fig:mmfa}. As a first overall result, we
note that the critical current is shifted to lower values due to the
instantaneous change of the effective potential for $\varphi_1$ by the fast
ramping of the current and by the quantum decay of $\varphi_2$, allowing the
heavy junction to transform potential into kinetic energy.

Another prominent change is in the form of the transition line, specifically,
the `cut' of the first tip of $j_c(k)$: The first segment at large values of
$k$, cf.\ Fig.\ \ref{fig:mmfa}, corresponds to the conventional case for a
$^+$-type line; the criterion for delocalization involves the full conversion
of potential to kinetic energy of the heavy junction and the subsequent
tunneling of the quantum junction right at the turning point of the classical
junction, cf.\ inset (1) of Fig.\ \ref{fig:mmfa}. Following the
$j_c^+(k)$-line further with decreasing $j$, we reach a point where the
potential energy of the classical junction after the phase slip event is no
longer sufficient to overcome the next barrier and an additional increase in
$j$ is required to provide the missing energy. The segment (2) then involves
partial conversion of the initial potential energy within the two wells before
and after the phase slip event.  Finally, moving along the segment (2) towards
lower $k$ the value of $\varphi_1$ where the phase slip occurs decreases and
finally reaches $\varphi_1=0$ (cf.\ Fig.\ \ref{fig:mmfa}, inset (3)). The 
critical line then continues along the third segment which is of the typical
$^-$-type, with the quantum junction undergoing tunneling before the
classical junction starts moving. After tunneling, a large current is required
to lower the barrier in the cosine potential of the classical junction, cf.\
Fig.\ \ref{fig:mmfa}, inset (3). Note that the differentiation between positively sloped 
segments of type (1) and type (2) persists to segments determined by a
higher fluxon index $n$, although the difference is invisible on the scale of
Fig.\ \ref{fig:mmfa}.

%

\subsection{Moderate ratios of $C_1/C_2$: Probabilistic effects}
\label{sec:farthest}

Above, we have assumed the limit of a very large ratio $C_1/C_2$. In the
experiment, the dynamical asymmetry of the SQUID is implemented by shunting
one of the two junctions with an external capacitance.\cite{Steffen:2006} This
technique allows to reach large capacitance ratios of the order of
$C_1/C_2\lesssim 10^4$. Given this finite mass ratio, one may enter a new
dynamical regime where the classical junction is pulled out of its metastable
state more efficiently. This is realized when the quantum junction attains a
high energy state residing at the opposite side of the parabolic potential
(large values of $\varphi_2$ in Fig.\ \ref{fig:decay}) and holds on to it
while dragging the heavy junction out of its metastable state. This requires a
fast\cite{footnote4} current ramping $\partial_t j \gg 1/
\tilde\tau_2$ (on the scale of the intervalley motion of the light junction),
pushing the quantum junction to high potential energies during the ramp-up
process. Furthermore, the condition $\omega_{p,1} \gg 1/\tilde\tau_2$ has to
be satisfied to allow for motion of the heavy junction before relaxation of the
quantum junction (from a side-well local ground state) 
back to the global minimum. 

In order to simplify the situation, we assume a (weak) finite damping of the
classical junction, 
$\omega_{p,1}^{-1}\ll\tau_1 < \tilde\tau_2$, assuring that the classical
junction relaxes before tunneling of the quantum junction (this assumption is
merely a convenience, allowing us to ignore the availability of kinetic energy
for the classical junction upon fast ramping).  We then start from the local
ground state of the original potential well at $\varphi_1=\varphi_2=\arcsin
j$.  The decay proceeds in the direction of $\varphi_2$ in the effective
potential $v_\mathrm{eff} [\arcsin j](\varphi_2)$ and can end up in any of the
lower local ground states for the quantum junction. We denote the energy of
the local ground state of the quantum junction in the $n$-th well by
$\tilde\varepsilon_0^n$ (not to be confused with $\varepsilon_n$, the $n$-th
excited state in the spectrum of the quantum junction). We estimate these
energies as
\begin{equation}
   \tilde\varepsilon_0^n(\varphi_1)
   \approx v(\varphi_1,\varphi_2^{\mathrm{min},n}),
\label{eq:e0n}
\end{equation}
where $\varphi_2^{\mathrm{min},n}(\varphi_1)$ is the solution to Eq.\
\eqref{eq:min_2} near $\varphi_2\approx 2 \pi n$ at fixed $\varphi_1$ and we
have neglected the zero point energy $\hbar\omega_{p,2}/2$.

\begin{figure}[htb]
\begin{center}
\includegraphics{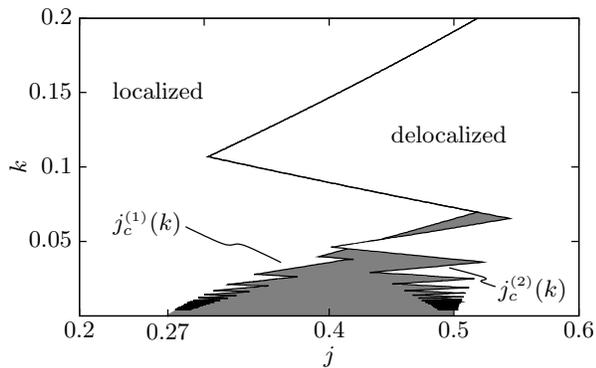}
\caption{Effective critical currents of the dynamically asymmetric SQUID for
$\tilde \tau_2\gg \omega_{p1}^{-1}$ and fast ramping. The system always turns
resistive for $j>j_{c}^{\scriptscriptstyle (2)}(k)$ but never turns resistive
for $j<j_{c}^{\scriptscriptstyle (1)}(k)$. For $j_{c}^{\scriptscriptstyle
(1)}<j<j_{c}^{\scriptscriptstyle (2)}$ the delocalization of the system is
determined by the statistical nature of the decay process of the quantum
junction involving the side minima.}
\label{fig:farthest}
\end{center}
\end{figure}

The dynamics of $\varphi_1$ depends strongly on the fluxon index $n$ of the
state of the quantum junction after the decay: the larger $n$, the stronger is
the force on $\varphi_1$ facilitating its escape. We
can define two extreme values for the critical current, a lower limit
$j_{c}^{\scriptscriptstyle (1)}(k)$ arising from the largest attainable fluxon
index $n$, and an upper limit $j_{c}^{\scriptscriptstyle (2)}(k)$ associated 
with the index $n$ of the global minimum of the quantum junction. 
We estimate the maximal index $n$ by comparing
the energy of the state of the quantum junction before its inital decay with
the height of the barrier blocking the access to the $n$-th minimum; the
numerical solution of the relation
\begin{equation}
v_\mathrm{eff}[\varphi_1=\arcsin j](\varphi_2=\arcsin j)=v_\mathrm{eff}
(\varphi_2^{\mathrm{max},n}), 
\label{eq:farthest}
\end{equation}
with $\varphi_2^{\mathrm{max},n}$ the solution to Eq.\ \eqref{eq:min_2} near
$\varphi_2^\mathrm{max}\approx (2 n-1) \pi$ at $\varphi_1=\arcsin j$ provides
us with the line $j_{c,n}^{{\scriptscriptstyle (1)}+}(k)$. The subsequent
stability analysis of $\varphi_1$ generates the lines
$j_{c,n}^{{\scriptscriptstyle (1)}-}(k)$; their combination into
$j_{c}^{\scriptscriptstyle (1)}(k)$ is displayed in Fig.\ \ref{fig:farthest}.

 The shape of $j_{c}^{\scriptscriptstyle (2)}(k)$ is found as described before:
We determine the index $n$ of the global minimum of $v_\mathrm{eff}
(\varphi_2)$ and deduce the motion of $\varphi_1$ in the effective potential
$\tilde\varepsilon_0^n(\varphi_1)$. The reduction in critical current due to the 
massive dynamics of $\varphi_1$ is not as pronounced as in
Fig.\ \ref{fig:mmfa}, since the maximal critical current $j_{c}^{\scriptscriptstyle
(2)}(k)$ involves the process where the classical junction dissipates most of
its kinetic energy: the quantum junction decays to the $(n-1)$ fluxon state,
the classical junction relaxes to its local minimum near $\varphi_1 \gtrsim
\arcsin j$, whereupon the quantum junction decays further to its global ground
state with fluxon index $n$. Only the potential energy gained by the classical junction 
in the last step can be transformed to kinetic energy, making the resulting line
 shown in Fig.\ \ref{fig:farthest} differ 
 from $j_c(k)$ in the overdamped case (Fig.\ \ref{fig:phase_diagram_vv}).

The two critical lines $j_{c}^{\scriptscriptstyle (1)}(k)$ and
$j_{c}^{\scriptscriptstyle (2)}(k)$ define a broad intermediate regime, see
the grey area in Fig.\ \ref{fig:farthest}, where the decay of the system is of
probabilistic nature: whereas no M\"unchhausen decay is possible for
$j<j_c^{\scriptscriptstyle (1)}(k)$, the decay of the system in the
intermediate regime depends on which minimum the quantum junction $\varphi_2$
decays to. For $j> j_{c}^{\scriptscriptstyle (2)}(k)$ the system is always
bound to decay.

The critical line $j_{c}^{\scriptscriptstyle (2)}(k)$ crosses twice the
critical line  $j_{c}^{\scriptscriptstyle (1)}(k)$ near $k\approx 0.5$ and for
$n=2$. This special situation arises when the most distant accessible minimum
and the global minimum are neighbors (with indices $n=2$ and $n=1$) and hence
strongly interrelated.  With decreasing $k$, both the height of the maximum
near $\varphi_2\approx 3\pi$ (determining $j_c^{\scriptscriptstyle (1)}(k)$)
and the position of the minimum at $\varphi_2\approx 4\pi$  ($n=2$) with respect
to the one at $\varphi_2\approx 2\pi$ ($n=1$; the competition between these
minima defines $j_c^{\scriptscriptstyle (2)}(k)$) are lowered. The latter
effect is slightly stronger, leading to the crossing of the lines under a
small angle.  Upon further lowering of $k$, the relevant fluxon index for
$j_c^{\scriptscriptstyle (1)}(k)$ increases twice as fast as for
$j_c^{\scriptscriptstyle (2)}(k)$, preventing any further crossings of the
critcal lines.

To guarantee reasonable time scales in the experiment requires a fast quantum
dynamics near the global minimum, hence the ratio $E_J/\hbar\omega_{p,2}$
should be small, of order unity. For a fast ramping, this is in conflict with
the requirement of large $\tilde\tau_2$ in all side-wells; as a result, not
all reachable side minima may be relevant for the dynamics of $\varphi_1$
(i.e., $\tilde \tau_2\gg \omega_{p,1}^{-1}$ does not hold), effectively
shifting the lower critical current $j_c^{\scriptscriptstyle (1)}(k)$ to
larger values.

So far we have neglected any quantum corrections for the variable $\varphi_1$;
however, with a finite ratio $C_1/C_2$ such corrections may become
relevant. Modifications due to quantum fluctuations in the heavy
degree of freedom are twofold: On the one hand, we have disregarded the
tunneling of $\varphi_1$ through a small residual barrier, leading to a
broadening of all lines of the $^-$-type. On the other hand, we have assumed
the dynamics of $\varphi_2$ to occur in the effective potential
$v_\mathrm{eff} [\varphi_1](\varphi_2)$, with $\varphi_1$ fixed, while for a
finite value of $C_1$ the ground state wave function of the heavy junction
acquires a finite width of the order of $\delta\varphi_1\sim
(8E_{c1}/E_J)^{1/4}$. This leads to a broadening of all lines of the
$^+$-type, which can be interpreted as $\varphi_1$ taking part in the
tunneling of $\varphi_2$.

\section{Experimental Implementation}
\label{sec:remarks}

The running state following a M\"unchhausen decay is associated with a finite
time-averaged voltage across the device. The above results thus can be tested
experimentally by measuring the voltage drop as a function of the applied bias
current $J$ and of the inductance $L$ of the SQUID.

Mapping any part of $j_c(k)$ requires changing the coupling $k\propto L^{-1}$.
The latter is determined by the total inductance of the loop $L =
L_\mathrm{geo} + L_\mathrm{kin}$, comprising the geometric ($L_\mathrm{geo}$)
and kinetic ($L_\mathrm{kin}$) inductances of the loop. The former is of the
order of the loop perimeter and accounts for the magnetic field energy. The
kinetic inductance relates to the geometric quantity via $L_\mathrm{kin} \sim
L_\mathrm{geo} \lambda^2/r^2$, with $r$ the radius of the wire and $\lambda$
denoting the superconducting penetration depth. Usually, the inductance is
dominated by its geometric part. However, a large inductance, as desired for
our setup, can be installed by using a thin wire.  Furthermore, the inductance
is fixed after fabrication and its modification (e.g., via installing a
diamagnetic shield) is rather difficult.  Another way to trace the critical
current line is obtained by applying an external magnetic flux $\Phi_e$ to the
sample, in which case the potential Eq.\ \eqref{eq:v} has to be replaced by
\begin{multline}
v(\varphi_1,\varphi_2)=1-\cos\varphi_1+1-\cos\varphi_2\\ 
-j(\varphi_1+\varphi_2)+\frac{k}{2}(\varphi_1-\varphi_2-2\pi\Phi_e/\Phi_0)^2.
\label{eq:v_phi_e}
\end{multline}
The critical current $j_c$ then can be studied as a function of $\Phi_e$. The
analysis proceeds in the same way as before: a change in $k$
altering the opening angle of the parabola in $v(\varphi_1,\varphi_2)$ is
replaced by a shift in the parabola's position due to the applied flux
$\Phi_e$.
\begin{figure}[htb] 
\begin{center} 
\includegraphics{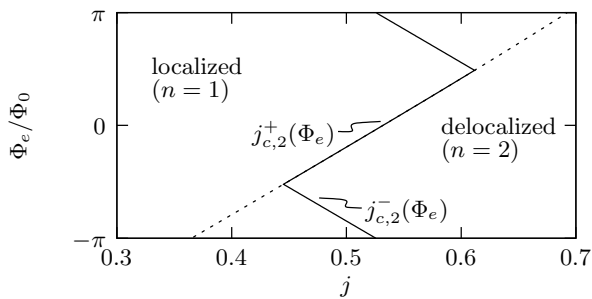} 
\caption{Phase diagram of the dynamically asymmetric SQUID in the strong
damping case as a function of the externally applied flux $\Phi_e$ and bias
current $j$. Here, the coupling $k \propto L^{-1}$ is fixed at a value 
$k=0.06$; the critical lines in the diagram correspond to $n=2$. The
critical line is periodic in $\Phi_e$ with a period $\Phi_0$, a consequence of
the invariance under the replacements $\Phi_e/\Phi_0 \rightarrow
\Phi_e/\Phi_0\pm 1$ and $n\rightarrow (n \pm1)$.}
\label{fig:pe} 
\end{center} 
\end{figure}

As an illustration, we discuss here the most simple case of an overdamped
dynamics, analogous to Sec.\ \ref{sec:overdamped}. In a first step, the
system relaxes to the ground state at the minimum of the potential
$v(\varphi_1,\varphi_2)$, Eq.\ \eqref{eq:v_phi_e}, as given by the solutions
$\varphi_1^{\mathrm{min}}, \varphi_2^{\mathrm{min}}$ of
\begin{eqnarray}
   \sin(\varphi_1^{\mathrm{min}})-j+k\left(\varphi_1^{\mathrm{min}}
   -\varphi_2^{\mathrm{min}}-2\pi\frac{\Phi_e}{\Phi_0}\right)\!&=&\!0,
   \label{eq:min_phi_e_1}\\
  \sin(\varphi_2^{\mathrm{min}})
   -j-k\left(\varphi_1^{\mathrm{min}}
   -\varphi_2^{\mathrm{min}}-2\pi\frac{\Phi_e}{\Phi_0}\right)\!&=&\!0,
   \label{eq:min_phi_e_2}
 \end{eqnarray}
near $\varphi_1^{\mathrm{min}}\approx\varphi_2^{\mathrm{min}}\approx 0$. The
critical line $j^+_{c,n}(\Phi_e)$ then derives from the condition
\begin{equation}
k\approx\frac{j}{(2n-1)\pi-\varphi_1^{\mathrm{min}}(\Phi_e)+2\pi\Phi_e/\Phi_0},
\label{eq:glob_min_pe}
\end{equation}
where both the shift of the parabola $v_\mathrm{eff}(\varphi_2)$ by $2\pi
\Phi_e/ \Phi_0$ and the flux dependence of $\varphi_1^{\mathrm{min}}$ are
accounted for.  Similarly, Eq.\ \eqref{eq:approximate_critical_k} for
$k_{c,n}^-(j)$ is changed to
\begin{equation}
   k\approx\frac{1-j}{(2n-1/2)\pi+\arcsin(2j-1)+2\pi\Phi_e/\Phi_0},
   \label{eq:approx_crit_k_pe}
\end{equation}
from which we find the critical lines $j^-_{c,n}(\Phi_e)$.  The results of Eqs.\
\eqref{eq:min_phi_e_1} - \eqref{eq:approx_crit_k_pe} are invariant under the
shifts $\Phi_e/\Phi_0 \rightarrow \Phi_e/\Phi_0\pm 1$ and $n \rightarrow
(n\pm1)$. The M\"unchhausen decay thus can be studied within a finite
interval, e.g. $\Phi_e/ \Phi_0\in[-0.5,0.5]$, and the resulting phase diagram
is displayed in Fig.\ \ref{fig:pe}.

Biasing the SQUID with an external magnetic flux has a convenient side effect:
For negative flux $\Phi_e$, the crossing point of $j_{c,n}^+(k)$ and
$j_{c,n}^-(k)$ is shifted to larger values of $k$ and the tunneling barrier
for $\varphi_2$ is lowered, allowing to study the M\"unchhausen effect using a
SQUID with considerably smaller inductance and hence smaller size.

\section{Conclusion}
\label{sec:conclusion}

We have shown how a system consisting of a `quantum' degree of freedom coupled
to a `classical' one, implemented with a two-junction SQUID, can perform a
complex decay process out of a zero-voltage state involving quantum tunneling
of the `light' junction and classical motion of the `heavy' junction. We have
analyzed the cases of strong and weak damping, for different methods of
preparation (fast and slow ramping), and for various ratios of the
capacitances $C_1/C_2$ and have found the effective critical current $j_c(k)$
exhibiting a common characteristic shape. Its zig-zag like structure
originates in two competing effects: while the number $n$ of flux units which
can enter the SQUID is increased with decreasing $k$, the current redirected
over the heavy junction at given $n$ is decreased. For junctions with equal
critical currents, the critical line approaches the value $j_c(k \rightarrow
0) \rightarrow 1/2$ in the dissipative case; asymmetries in the critical
currents of the two junctions lead to a shift of the critical line $j_c(k)$ to
values straddling the critical current of the heavy junction.  Going over from
dissipative to massive dynamics, the kinetic energy stored in the system
allows to better overcome the potential barriers and the critical line shifts
to smaller values of $j$; furthermore, hysteretic effects, similar to the
situation encountered for underdamped junctions, and even probabilistic
behavior may show up.

\begin{acknowledgments}
We thank A.\ Larkin, A.\ Ustinov, G.\ Lesovik, A.\ Lebedev, A.\ Wallraff, and
E.\ Zeldov for interesting discussions and acknowledge support of the Fonds
National Suisse through MaNEP.
\end{acknowledgments}


\begin{thebibliography}{10}
\bibitem{footnote1}
We use modified Gaussian units, 
where $C$ and $L$ are measured 
in units of length.
\bibitem{footnote3}
While the correction $E_0$ might not be 
immaterial, its dependence on $\varphi_1$ 
is small.
\bibitem{footnote4}
When the ramping is slow, the modifications 
with respect to the discussion in 
Sec.\ \ref{sec:adiabatic} are negligible: Due 
to the slow ramping, the quantum junction 
remains always in its ground state and hence the
classical junction moves adiabatically in the 
effective potential $\varepsilon_0(\varphi_1)$ 
created by the quantum junction.  The resulting
critical current $j_c(k)$ is identical to the one 
presented in Fig.\ \ref{fig:phase_diagram_adiabatic}.
\end{thebibliography}
\end{document}